\documentclass[sigconf]{acmart}

\usepackage{csquotes}
\makeatletter
\@ifclasswith{acmart}{anonymous}
{%
    \usepackage[true]{anonymous-acm}
}{%
    \usepackage[false]{anonymous-acm} 
}
\makeatother
\usepackage{ifdraft}
\usepackage{etoolbox} % for global bool
\usepackage{xspace}
\usepackage{color}
\usepackage{tabularray}
\usepackage{comment}
\usepackage{cleveref}
\renewcommand*{\cref}{\Cref}
\usepackage{hyphenat}
\usepackage{array}
\usepackage{booktabs}
\usepackage{tabularx} 
\usepackage{multirow}
\usepackage{longtable}
\usepackage{subfigure}
\usepackage{soul}
\usepackage{xcolor}
\usepackage{utfsym}%MnSymbol,utfsym,arev
\definecolor{lightyellow}{RGB}{255, 251, 204}
\usepackage{acmart-taps}
\usepackage{graphicx}
\usepackage{subcaption}

\AtBeginDocument{%
  \providecommand\BibTeX{{%
    \normalfont B\kern-0.5em{\scshape i\kern-0.25em b}\kern-0.8em\TeX}}}

\usepackage{longtable}
\usepackage{colortbl}
\usepackage{multirow}
\usepackage{subcaption}
\usepackage{tabularray}
\usepackage{csquotes}

\newcommand{\elide}[1]{\textelp{}} % produces ... instead of argument
 
\newcommand{\textcite}[1]{\citeauthor{#1}~\cite{#1}}

\usepackage{tikz}

\usepackage[textsize=scriptsize]{todonotes}
\definecolor{lightyellow}{RGB}{255, 251, 204}
\definecolor{lightgreen}{RGB}{204, 255, 204}

\AtBeginDocument{%
  \providecommand\BibTeX{{%
    Bib\TeX}}}

\copyrightyear{2025}
\acmYear{2025}
\setcopyright{rightsretained}
\acmConference[CHI EA '25]{Extended Abstracts of the CHI Conference on Human Factors in Computing Systems}{April 26-May 1, 2025}{Yokohama, Japan}
\acmBooktitle{Extended Abstracts of the CHI Conference on Human Factors in Computing Systems (CHI EA '25), April 26-May 1, 2025, Yokohama, Japan}\acmDOI{10.1145/3706599.3719821}
\acmISBN{979-8-4007-1395-8/2025/04}

\begin{document}

\title{Can Large Language Models Grasp Concepts in Visual Content? A Case Study on YouTube Shorts about Depression}

\author{Jiaying ``Lizzy'' Liu}
\authornote{Both authors contributed equally to this research.}
\orcid{0000-0002-5398-1485}
\affiliation{%%
  \institution{School of Information \\ The University of Texas at Austin}
  \city{Austin}
  \state{Texas}
  \country{USA}
 }
\email{jiayingliu@utexas.edu}

\author{Yiheng Su}
\authornotemark[1]
\orcid{0009-0001-8021-429X}
\affiliation{%
  \institution{Artificial Intelligence and Human-Centered Computing (AI\&HCC) Lab \\ The University of Texas at Austin}
  \city{Austin}
  \state{Texas}
  \country{USA}
  }
\email{sam.su@utexas.edu}

\author{Praneel Seth}
\orcid{0009-0001-4665-0776}
\affiliation{%
  \institution{Computer Science Department\\ The University of Texas at Austin}
  \city{Austin}
  \state{Texas}
  \country{USA}
}
\email{praneelseth@utexas.edu}

\renewcommand{\shortauthors}{Liu et al.}
\renewcommand{\shorttitle}{Can Large Language Models Grasp Concepts in Visual Content? \\A Case Study on YouTube Shorts about Depression.}
\begin{abstract}
Large language models (LLMs) are increasingly used to assist computational social science research. While prior efforts have focused on text, the potential of leveraging multimodal LLMs (MLLMs) for online video studies remains underexplored. We conduct one of the first case studies on MLLM-assisted video content analysis, comparing AI's interpretations to human understanding of abstract concepts. 
We leverage LLaVA-1.6 Mistral 7B to interpret four abstract concepts regarding video-mediated self-disclosure, analyzing 725 keyframes from 142 depression-related YouTube short videos. We perform a qualitative analysis of MLLM's self-generated explanations and found that the degree of operationalization can influence MLLM's interpretations. Interestingly, greater detail does not necessarily increase human-AI alignment. We also identify other factors affecting AI alignment with human understanding, such as concept complexity and versatility of video genres. Our exploratory study highlights the need to customize prompts for specific concepts and calls for researchers to incorporate more human-centered evaluations when working with AI systems in a multimodal context.
\end{abstract}




\maketitle
\section{Introduction}
Video-sharing platforms such as YouTube \cite{liu_modeling_2024}, TikTok \cite{schaadhardt_laughing_2023}, and Instagram \cite{andalibi_self-disclosure_2017} are rich data sources for research in human-computer interaction and computational social sciences. 
However, traditional methods for analyzing videos, like digital ethnography \cite{kubitschko_innovative_2016} and content analysis \cite{drisko_content_2016}, are labor-intensive with limited scalability \cite{10.1145/3544548.3581107}.
% Given the ubiquity of video content in social media, 
Consequently, there is a rising demand for automated approaches to analyze multimodal (visual, textual, audio) content \cite{bartolome2023literature}.

One successful strategy is leveraging LLMs to augment text-based content analysis, improving open coding efficiency \cite{chew_llm-assisted_2023} and enabling collaborative coding frameworks \cite{gao_collabcoder_2023, xiao_supporting_2023}. Emerging Multimodal LLMs (MLLMs) like LLaVA \cite{liu2024improved} and GPT-4 \cite{openai2024gpt4} demonstrate promise for understanding visual information at scale \cite{wang2024comprehensive}. However, few works have investigated how MLLMs can best assist content analysis of videos \cite{tang2023video, Yin2024MLLMSurvey}. Preliminary work \cite{manikonda_modeling_2017} suggests that MLLMs may struggle to capture abstract visual concepts, such as video presentation style \cite{liu_harnessing_2024}, limiting their applications beyond objective entity or action recognition in video analysis \cite{NEURIPS2023_4f8e27f6, Chang2024EvaluationLLM, alsagheer_evaluating_2024}. 

This case study thus aims to explore the capability of MLLMs to understand abstract concepts in multimodal contexts. Specifically, we investigate how LLaVA-1.6 Mistral 7B interprets four concepts related to depression and self-disclosure behaviors in short YouTube videos, assessing the MLLM's alignment with human understanding. We aim to explore:

    % How to prompt MLLM to assist social concept annotation in video content analysis?
\begin{itemize}
    \item[\textbf{RQ1:}] \textbf{How can social concepts be operationalized to guide MLLMs in interpreting video content?}
    \item[\textbf{RQ2:}] \textbf{What factors affect MLLM's alignment with human interpretations of social concepts in videos?}
\end{itemize}

Echoing the emerging trend of LLM-assisted content analysis, our case study is one of the earliest efforts to leverage MLLMs for video content analysis: 1) We experiment with harnessing an MLLM for annotating abstract visual concepts with structured and explainable outputs; 2) We examine the MLLM's explanations and reveal contextual factors that affect MLLM's alignment with human understanding of abstract social concepts. 3) We discuss implications for designing robust, human-centered workflows for future MLLM-assisted video content analysis.

\section{Context: Mental Health Disclosure on Video-Based Social Media}

Individuals increasingly use digital platforms to share their mental health experiences and seek support online \cite{feuston_everyday_2019}. 
While prior research has extensively focused on text-based platforms like Twitter \cite{choudhury_mental_2014} and Reddit \cite{pendse_marginalization_2023}, visual-based platforms like Instagram \cite{andalibi_sensitive_2017} and YouTube \cite{huh_health_2014, liu_health_2013} are growing in popularity for self-disclosure documentation. 
% and personal experience documentation. 
% individuals are now increasingly utilizing visual tools, such as Instagram photos \cite{andalibi_sensitive_2017} and YouTube vlogs \cite{huh_health_2014, liu_health_2013}, to document their illness journeys.

Visual content offers unique self-disclosure opportunities distinct from textual modalities \cite{manikonda_modeling_2017}. Specific image genres like selfies, social relationships, and captioned images can convey emotional distress, calls for help, and vulnerability in powerful ways \cite{andalibi_self-disclosure_2017}. Similar to the influence of linguistic features on engagement for text-based social media posts, prior studies have highlighted the significant role of visual representations in shaping viewer perception \cite{li_understanding_2024} and supportive behaviors (e.g., comments) \cite{hu_making_2023}. However, which features of the visual representations and how they influence viewer engagement remain unclear. % The multimodal nature of video may enable deeper disclosure and stronger creator-audience connections \cite{liu_modeling_2023, liu_understanding_2024}.

This work thus aims to extend prior text-based online health communication research into the underexplored video-based social media, where self-disclosure may be communicated through interactive language and visual cues. Specifically, we investigate how visual features moderate the relationship between self-disclosure and video engagement (e.g., likes and comments) in depression-related YouTube shorts. Addressing this question can reveal insights such as identifying visual markers of distress, rhetorical framing of health narratives, and emergent phenomena in visually diverse content to inform the design of more supportive communities on video-sharing platforms. Given the challenges of manual annotation for large-scale video content analysis, we leverage MLLMs for assistance.

We selected four concepts (Table \ref{tab:concepts}) that shape video-mediated self-disclosure. Presenting and interacting styles represent distinct approaches to structuring and delivering video narratives, which influence audience engagement \cite{lachmar_mydepressionlookslike_2017, andalibi_self-disclosure_2017}. Visual diversity and arousal are unique for video-based communication, influencing viewers' attention and perception of content engagement \cite{metallinos2013television, seckler2015linking}. These visual characteristics are indicative cues to determine how effectively mental health content resonates with and engages viewers.

\aptLtoX[graphic=no,type=html]{\begin{table*}[h]
\caption{Four abstract visual concepts shaping video-mediated self-disclosure.}
\label{tab:concepts}
\footnotesize
\begin{tabular}{|l|c|}
\hline
\cellcolor[HTML]{dedede}{Concept} & \cellcolor[HTML]{dedede}{Definition} \\
\hline
Presenting & Presenting style involves the delivery of information, typically accompanied by visual aids like slides or graphics \cite{kelly-hedrick_its_2018}. \\
\hline
Interacting & Interacting refers to creators establishing a simulated interpersonal relationship with their audience, fostering a sense of engagement and connection \cite{kelly-hedrick_its_2018}. \\
\hline
Diversity & Diversity of an image includes varied scenes, color variation, compositional complexity, and originality of the image \cite{seckler2015linking}. \\
\hline
Arousal & Arousal refers to the degree of alertness or excitement elicited by the stimulus such as dynamic visual elements and emotional intensity \cite{metallinos2013television}. \\
\hline
\end{tabular}
\end{table*}}{\begin{table*}[h]
\caption{Four abstract visual concepts shaping video-mediated self-disclosure.}
\label{tab:concepts}
\footnotesize
\begin{tblr}{
width = \textwidth,
colspec = {|Q[l,0.1]|Q[l,0.8]|},
hlines,
row{1} = {font=\bfseries,c,gray9},
}

Concept & Definition \\

Presenting & Presenting style involves the delivery of information, typically accompanied by visual aids like slides or graphics \cite{kelly-hedrick_its_2018}. \\
%& Does this picture communicate in a presenting style, yes or no? Explain your answer. 

Interacting & Interacting refers to creators establishing a simulated interpersonal relationship with their audience, fostering a sense of engagement and connection \cite{kelly-hedrick_its_2018}. \\
%& Does this picture portray an interacting style, yes or no? Explain your answer. 

Diversity & Diversity of an image includes varied scenes, color variation, compositional complexity, and originality of the image \cite{seckler2015linking}. \\
%& What level of diversity does this image communicate, low or high? Explain your answer. \\

%Emotional Valence & Emotional valence is the value associated with a stimulus as expressed on a continuum from pleasant to unpleasant or from attractive to aversive. & Is the emotional valence of the picture positive, negative, or neutral? Explain your answer. \\

Arousal & Arousal refers to the degree of alertness or excitement elicited by the stimulus such as dynamic visual elements and emotional intensity \cite{metallinos2013television}. \\
%& What level of arousal does this image communicate, low or high? Explain your answer. \\

\end{tblr}
\end{table*}}

\section{Methodology}

\subsection{Dataset}

Using the query "depression" with the YouTube Data API, we collected the metadata (e.g., title, channel, duration) of 3,892 videos uploaded by February 2024.  %we randomly sampled 10\% of the videos (391 videos), resulting in a final set of 244 videos. 
%Since current MLLMs that directly process videos perform worse than image-based ones, we follow \cite{liu_harnessing_2024} by extracting and annotating keyframes in videos instead. Due to computational constraints, 
We randomly selected 150 videos and downloaded them using \texttt{YoutubeDownloader}\footnote{\url{https://github.com/Tyrrrz/YoutubeDownloader}}. Following Liu et al. (2024) \cite{liu_harnessing_2024}, due to computational constraints and the current MLLM's limited context window to process videos \cite{he2024malmm}, we applied FFmpeg \cite{tomar2006converting} to extract representative keyframes in videos. 
FFmpeg is a standard video processing method for identifying key moments in videos \cite{Mahasseni_2017_CVPR, HONG2018611, Lee_2023_CVPR}. Here, we employed FFmpeg to extract frames where the structural similarity index (SSIM) \cite{wang2004image} difference exceeded 0.3, ensuring the selection of visually distinct frames. Additionally, we filtered out low-quality frames (e.g., transitional frames, blurry, black screens) through manual inspection and obtained 725 keyframes across 142 videos.

Our study qualifies for exemption under our Institutional Review Board guidelines. Nevertheless, recognizing the sensitive nature of mental health topics, we safeguard video creators' privacy by anonymizing their identities through obscuring facial features. Further discussion of ethical considerations can be found in Appendix~\ref{sec:ethics}.

\begin{figure}[h]
    \centering
    \includegraphics[width=\linewidth]{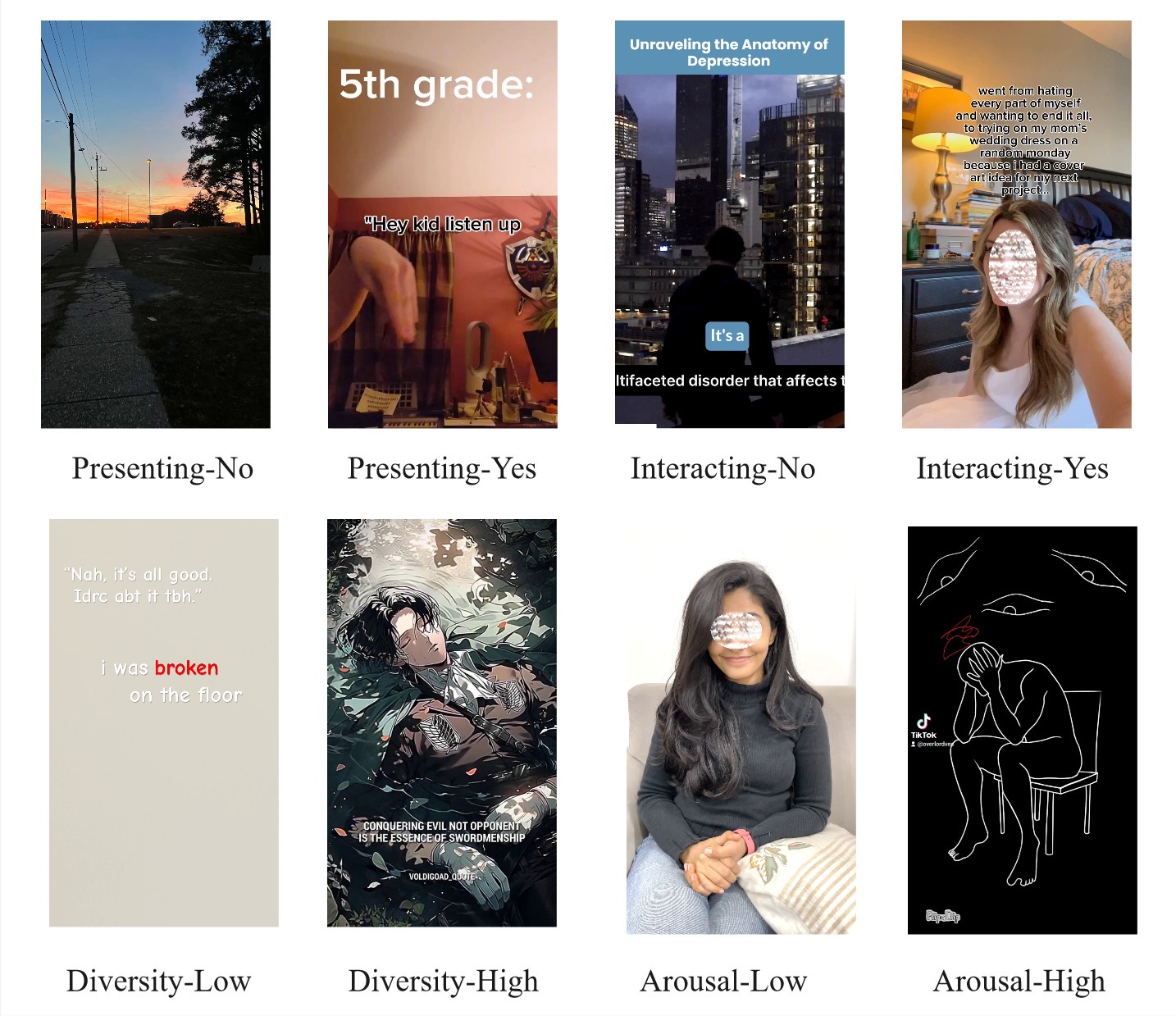}
    \caption{Examples of human interpretations of the four selected concepts. We annotate Yes/No for \textit{presenting} and \textit{interacting}, High/Low for \textit{diversity} and \textit{arousal}. We then compare human interpretations with the MLLM interpretations to evaluate human-AI alignment.}
    \Description{Eight examples of human annotations of the four selected concepts.}
    \label{fig:conceptexamples}
\end{figure}

\subsection{MLLM Concept Annotation: Models and Prompts}

We first tested Video-LLaMA \cite{zhang-etal-2023-video} on 10 sample videos and observed significant challenges in concept comprehension and high computational costs. This observation aligns with prior findings that Video LLMs generally underperform compared to Image LLMs \cite{liu_tempcompass_2024}. Given these limitations, we instead selected \texttt{llava-v1.6-mistral-7b-hf}\footnote{\url{https://huggingface.co/llava-hf/llava-v1.6-mistral-7b-hf}} \cite{Liu_2024_CVPR} to analyze keyframes and will henceforth refer to this model as (the) MLLM for convenience.

To investigate the MLLM's comprehension of abstract visual concepts (Table \ref{tab:concepts}), operationalizing these concepts is essential for articulating them effectively. To address RQ1 and explore how to operationalize the concepts for MLLM prompt configuration, we tested four strategies and evaluated their effectiveness. Specifically, we implemented four prompting configurations with progressively increasing levels of operational guidance to strike a balance between clarity and flexibility. See Appendix~\ref{sec:LLaVAPrompts} for all prompt configurations.

\begin{itemize}
    \item \textbf{Naive}: The MLLM is directly queried for the presence or extent of the concept without any additional contexts.
    \item \textbf{Simple}: A short definition is added to the naive query.
    \item \textbf{Detailed}: A detailed definition with three abstract manifestations is added to the naive query.
    \item \textbf{Open-minded}: Similar to the detailed prompt, but also explicitly encourages the MLLM to consider other scenarios not already stated. 
\end{itemize}

Established practices in prompt engineering inform our prompting configurations. For instance, the \textbf{Detailed} configuration aligns with in-context learning by incorporating prototypical examples to serve as implicit "demonstrations" \cite{min-etal-2022-rethinking}. The \textbf{Open-minded} configuration is inspired by chain-of-thought (CoT), a technique that aims to improve LLM logical reasoning by incorporating directives like "think step-by-step" in prompts \cite{NEURIPS2022_8bb0d291}. 
In our context, we aim to mitigate potential constraints introduced by fixed definitions in \textbf{Simple} and \textbf{Detailed} configurations while balancing clarity and flexibility in how MLLMs interpret abstract concepts. Thus, we adapt CoT by explicitly instructing the model to "be open-minded" in the prompts to encourage divergent, reflexive thinking similar to tree-of-thought \cite{NEURIPS2023_271db992}. 
%While CoT typically promotes convergent thinking (to find the single best solution for a given problem) \cite{NEURIPS2023_45e15bae}, here we adapt CoT by employing a directive --- "be open-minded" --- to encourage divergent, reflexive thinking similar to tree-of-thought \cite{NEURIPS2023_271db992}.

We do not experiment with advanced configurations such as in-context learning or fine-tuning \cite{dong-etal-2024-survey, han2024parameter}, as we are interested in assessing the MLLM's off-the-shelf capabilities.

We tasked the MLLM with annotating each keyframe across four concepts: Yes/No for \textit{interacting} and \textit{presenting}, and High/Low for \textit{arousal} and \textit{diversity}. To ensure consistency, we prompted the MLLM to provide both interpretations and explanations simultaneously, reducing the likelihood of generating contradictory or hallucinated explanations. Keyframes were queried in temporal order for each video, while the order of prompt configurations and associated concepts were randomized per keyframe to mitigate potential biases.
Occasionally, the MLLM combines annotations (e.g., Yes/No) with explanations \cite{liu_harnessing_2024}. To isolate explicit annotations, we utilized \texttt{Llama-3.1-8B-Instruct}\footnote{\url{https://huggingface.co/meta-llama/Meta-Llama-3.1-8B-Instruct}} to parse the MLLM's interpretations. Following this, we manually reviewed all extracted annotations to verify the accuracy of the parsing process.

\subsection{Human Annotation Process}
\label{sec:humanLabel}

To obtain human interpretations, two authors independently coded a random sample of 200 keyframes, with a third author providing an additional vote to resolve disagreements. Figure~\ref{fig:conceptexamples} illustrates examples of human interpretations. After discussing disputes in a group meeting and ensuring that Intercoder Reliability (ICR) \cite{o2020intercoder} is higher than 75\%, the three coders split the remaining keyframes and coded them separately.  We dropped ambiguous keyframes and low-quality images (e.g., transitional frames, blurry, black screens) from further analysis. Ultimately, we obtain 725 frames across 142 videos with human concept annotations.

\subsection{Data Analysis}
%We employ a mixed-method approach to address the proposed research questions, leveraging quantitative measures to answer RQ1 but adopting a qualitative approach for RQ2. 

\textbf{Quantitative Comparisons.} To compare the four prompt configurations, we quantify human-AI (mis)alignment as the consistency between a prompt-concept pair and the corresponding human annotations. 
% We then employ statistical procedures to assess how human-AI alignment differs across the four prompt approaches for each concept. 
We then employ the bootstrapping approach from \cite{berg-kirkpatrick-etal-2012-empirical} to assess how human-AI alignment differs across configurations per concept. Please see Appendix~\ref{sec:bootstrap} for details. We discuss quantitative comparisons in Section~\ref{sec:RQ1}.

\textbf{Qualitative Analysis.} To investigate the underlying factors behind human-AI (mis)alignments, we first curated a focused dataset of instances where the MLLM's annotations diverged when using different prompting configurations.
Two authors then independently conducted thematic analysis \cite{braun2006using} on the MLLM's explanations for these keyframes. They met weekly to discuss emerging themes and patterns in the data, resolve any coding discrepancies through detailed discussion, and iterate on the coding scheme to establish definitions for each thematic category.
The analysis focused on several key dimensions, including the nature and patterns of annotation changes, the MLLM's reasoning and justification for modifications, contextual factors that appeared to influence changes, and the relationship between prompting configuration and annotation stability.
We summarize recurring themes and patterns in Section~\ref{sec:RQ2}. 

\section{Findings}
    
\subsection{Quantative Evaluation of MLLM-Human Alignment}
\label{sec:RQ1}

% sam: thoughts to cut space 
% 1) Instead of graph, change to table instead for cleaner reference -- 16 cells, each cell is a interval 
% 2) Now sure about how table 2 works honestly -- maybe just cut since it doesn't feel like currently it's contributing anyway? 

\begin{figure}[htbp!]
  \centering
  \includegraphics[scale=0.6]{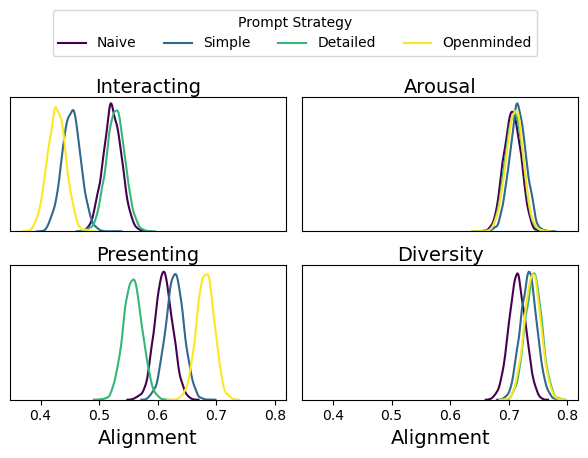}
  \caption{\label{fig:Density}
  Distribution of bootstrap alignment scores across prompt configurations and concepts. The MLLM demonstrates varying capabilities: no single prompt configuration consistently achieves the highest alignment across all concepts. 
  % outperforms others. 
  }
  \Description{Distribution of Bootstrap Alignment Scores Across Prompt Strategies Per Concept}
\end{figure}

Figure~\ref{fig:Density} shows the distribution of bootstrapped alignment scores across prompt configurations for each concept. The MLLM demonstrates varying capabilities: no single prompt configuration consistently achieves the highest alignment. 
% outperforms others.

The MLLM excels at abstract concepts like arousal and diversity but exhibits lower alignment and more variance for performative concepts like interacting and presenting. Under the naive approach, the MLLM performs well for concepts like interacting, arousal, and diversity, suggesting that the MLLM's prior knowledge of these concepts (derived from pre-trained data) aligns well with corresponding human conceptions. We only observe substantial alignment gains with more operationalization guidance for presenting. However, this effect is not monotonic (e.g., a more detailed prompt does not always lead to better alignment) and does not generalize to other concepts. Adding definitions may decrease alignment for presenting and interacting, restricting the MLLM's capabilities. We discuss the factors that impact annotations in Section~\ref{sec:RQ2}. 

\subsection{
Factors Affecting MLLM-Human (Mis)Alignment}
% RQ2: What factors affect the alignment between human and MLLM interpretations of social concepts in videos?
\label{sec:RQ2}

Evidently, concept operationalization is a key factor influencing human-AI alignment. By analyzing the MLLM's explanations, we offer qualitative insights into how and why operationalization impacts alignment. Additionally, we identify two further factors contributing to human-AI (mis)alignment: concept complexity and the diversity of genres.

\begin{figure*}[h]
    \centering
    \includegraphics[scale=1]{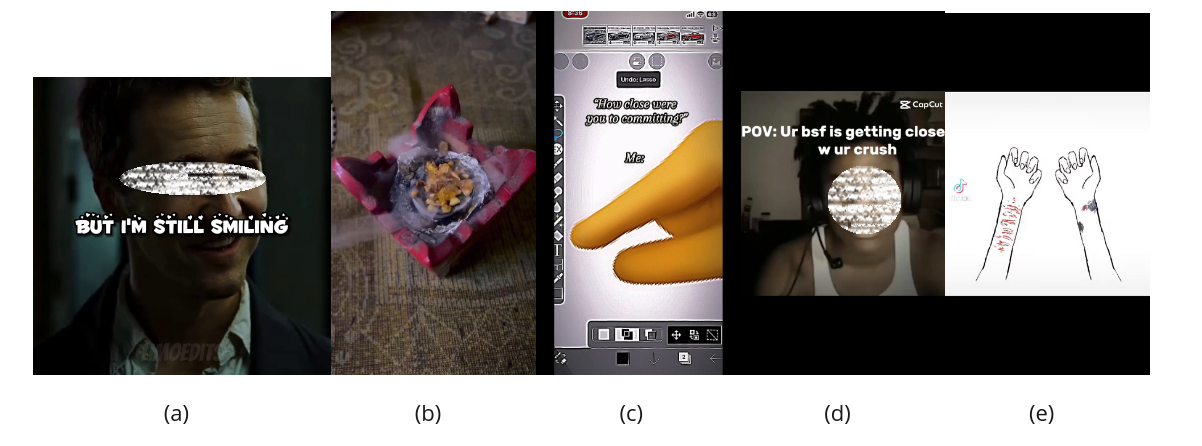}
    \caption{Problematic MLLM Annotations.}
    \Description{Five examples of keyframes that are discussed in the findings.}
    % \caption{Examples of Keyframes}
    \label{fig:examples}
\end{figure*}

\subsubsection{\textbf{Varying Concept Specification}}

Concept specification refers to the amount of detail in the prompts. For interacting and presenting (Figure~\ref{fig:Density}), auxiliary definitions may inadvertently prioritize "what is in the prompt" over the holistic context of the image, causing the MLLM to be less aligned with human perceptions. In contrast, the naive approach shows greater flexibility in capturing novel categories of presenting and interacting communication styles.

Figure \ref{fig:examples}-(a) illustrates the variability in the MLLM's interpretation of presenting style. 
When prompted naively, the MLLM correctly identifies (a) as presenting, stating that the superimposed caption is \textit{``a common technique used in presentations''}, complemented by \textit{``the person's facial expression, which appears to be a smile''}. 
% \textit{[the caption] is superimposed over the image of a person, which is a common technique used in presentations to convey a message or a quote...The person's facial expression, which appears to be a smile, complements the text and reinforces the message being presented.} 
Conversely, when prompted with simple or detailed configurations, the MLLM misclassifies (a), citing \textit{``no visible slide or graphic that would be associated with a presentation''} as evidence. This misclassification occurred because the detailed prompts explicitly exemplified presenting styles as "slides or graphics," limiting the MLLM from considering informal contexts of presenting style. In contrast, the openminded configuration correctly identifies (a), further underscoring that additional details can enhance clarity but reduce alignment if not carefully operationalized.

Without definitional constraints, the naive configuration can better capture nuanced social dynamics. In Figure \ref{fig:examples}-(b), the MLLM accurately described the interactive potential, noting the \textit{``dynamic and engaging''} style of the image to \textit{``[invite] the viewer to observe and possibly speculate about what is happening.''}

However, when prompted with a detailed configuration, the MLLM incorrectly claims that the image \textit{``is a still photograph''} with \textit{``no indication of a simulated interpersonal relationship or engagement with an audience''}.

We consistently observe this pattern of contradictory decisions for presenting and interacting queries, where explanations often highlight the absence of explicit elements outlined in the prompt. For example, keyframes without human presence or overt conversational styles (Figure~\ref{fig:examples}-(c)) were misclassified as non-interactive despite employing engaging nontraditional styles such as memes.

\subsubsection{\textbf{Varying Complexity of Concepts}} The complexity and scope of the four analyzed concepts vary, making some more challenging for the MLLM. For example, diversity is relatively straightforward, as it involves identifying and counting visual categories, a common pre-training task for MLLMs. Figure~\ref{fig:conceptexamples} illustrates this: the low-diversity image shows a plain background with simple text overlays, while the high-diversity image features a vibrant anime figure. Similarly, the MLLM effectively recognizes arousal levels through visual cues like facial expressions, body language, and visual intensity. In Figure~\ref{fig:conceptexamples}, the low-arousal image depicts a calm individual with relaxed features, while the high-arousal image shows an abstract figure with intense body language indicating distress.

In contrast, concepts like interacting and presenting are more challenging because they require situating visual cues within context. 
% Specifically, to understand what constitutes interacting or presenting, the MLLM must not only identify visual signals (e.g., captions, infographics) but also interpret them within the broader intentions of the author. 
For instance, in the "Presenting-Yes" image (Figure~\ref{fig:conceptexamples}), while the hand gesture 
% or the text overlay \pquote{5th grade: Hey kids listen up} 
might initially suggest interaction, the gesture is not directed at the audience but instead presents the scenario encoded in the text overlay (\textit{``5th grade: Hey kid listen up''}). In multimodal contexts, the meaning of one element (e.g., a visual cue) can influence, support, or contradict another (e.g., text). This demand to interpret co-dependent features holistically poses a novel challenge absent in text-only settings.

When MLLM's pre-trained knowledge diverges from human conceptions, naive queries often result in misalignment. 
% Querying annotations for these complex concepts naively may lead to a conception misalignment problem, where the MLLM's prior knowledge from pretraining diverges from human conceptions. 
We observe this quantitatively, as the Naive alignment for presenting is very low (Figure~\ref{fig:Density}). Qualitatively, in the "Presenting-Yes" image (Figure~\ref{fig:conceptexamples}), the MLLM incorrectly states that the image does not show presentation style, citing the absence of expected behaviors like \textit{``a speaker standing at a podium or a lectern''} and \textit{``a slide or a graphic''}. 
% \pquote{[the image does] not [show] a typical presentation style, which would typically involve a speaker standing at a podium or a lectern, with a slide or a graphic visible to the audience. Instead, this image appears to be a casual, candid moment captured in a room, likely a home or office setting, rather than a formal presentation environment.}. 
The MLLM fails to contextualize the informal setting and gesture as a valid presentation style, thus struggling to adapt to novel communicative contexts outside pretraining. Prompt engineering can help, as the MLLM correctly identifies this image for all other configurations besides naive.

\subsubsection{\textbf{Versatile Video Genres}}
\label{sec:Diverse}
The versatility of videos can challenge the MLLM's ability to understand social concepts. We identify two genres with relatively low alignment, highlighting the complexities of interpreting diverse content.

\paragraph{Mixture of textual and visual elements.}
Short videos often combine visuals with overlaying text, as shown in  Figure~\ref{fig:examples}-(a, c, d). When visual signals conflict with textual information, MLLMs (typically) prioritize textual over visual cues (since they were pre-trained with more text data), potentially leading to misinterpretations. For example, in Figure~\ref{fig:examples}-(d), the MLLM reasons that the image \textit{``does not directly portray an interacting style...as it is static''} but the text overlay \textit{``implies a narrative or a message that is meant to convey a sense of interaction.''} 
% : \pquote{[t]he image itself does not directly portray an interacting style, as it is a static image and does not depict any actions or interactions. However, the text overlay implies a narrative or a message that is meant to convey a sense of interaction.} 
Effectively synthesizing two potentially conflicting sources of information—visual and textual—is a unique and open challenge for MLLMs.

\paragraph{Non-human genres.}
Resonating Zhong et al. \cite{Zhong_2024_CVPRVLMMeme}, the MLLM struggles to interpret non-human video genres such as cartoons, memes, and abstract art, which often require cultural, emotional, or other contextual knowledge for accurate interpretations.
% These genres typically contain abstract content requiring cultural and other background knowledge for accurate interpretation.
For example, Figure \ref{fig:examples}-(e) depicts a hand-drawn image of self-harm behaviors, potentially signaling interaction intentions such as a call for help. 
% However, LLMs often cannot recognize implicit interaction cues in non-traditional visual formats. 
However, the MLLM failed to recognize implicit interaction cues and explained that \textit{``the drawing...does not exhibit any conversational language or behaviors that would suggest an interacting style''}. 

Fine-tuning or more sophisticated prompt engineering is likely needed to educate the MLLM on a broader range of visual storytelling techniques and cultural references.

\section{Discussion and Future Work}

We conduct an exploratory study with a single model, limited samples, and simple prompts, so our findings may not be generalizable. Computational constraints further prevented the inclusion of temporal context in videos, which may limit our findings. 
% other modalities like audio. 
Despite these limitations, our study offers insights into opportunities and challenges of leveraging Multimodal Large Language Models (MLLMs) to assist visual content analysis that are relevant regardless of the employed model. 
% for concept annotations of visual information. 
% While we acknowledge the limitations of our small sample as an exploratory study, we provide valuable insights and pave the way for more comprehensive evaluations in future research. 
Recognizing the inherent subjectivity of social concepts (even with high intercoder reliability), we use "alignment" rather than "accuracy" and contextualize our quantitative statistics with qualitative insights.
% though interpretations may still vary.
% We use \textbf{human-AI alignment} as a proxy to evaluate the feasibility of leveraging MLLMs for concept annotations. 
Our analysis illuminates key factors contributing to MLLM's misalignment from human understanding, including concept specifications, concept complexity, and versatility of video genres, which must be considered carefully when engineering prompts for MLLMs-assisted video content analysis.

\subsection{Harnessing MLLMs for Large-Scale Multimodal Content Analysis: Opportunities and Challenges}
 
MLLMs show potential in scaling visual content analysis. With appropriate operationalization, our results show that the MLLM can align highly with human perceptions, even for abstract concepts like presentation style. By expediting manual labeling, which is often time-intensive and costly \cite{drisko_content_2016}, MLLM can enable more comprehensive analyses of large datasets, potentially uncovering rare communication patterns that might otherwise go unnoticed in small-sample qualitative studies \cite{mcgrady2023dialing}. 
% high alignment for less complex concepts (diversity and arousal), which demonstrates the feasibility of leveraging MLLMs 
% opens up exciting opportunities for MLLM-assisted video content analysis at scale. MLLMs can expedite traditional manual labeling, often slow and costly \cite{drisko_content_2016}. 
% Notably, our study reveals that prompting MLLMs with naive queries can uncover novel communication styles that may differ from human perceptions, thus expanding the breadth of our understanding. 
Furthermore, MLLMs can enhance data quality by serving as a proxy for human intervention. In our pipeline, the MLLM accurately labeled low-quality frames as "Not Applicable," distinguishing them from frames that genuinely lacked the desired concept. This capability can help researchers filter noisy inputs by inspecting ambiguous model outputs and explanations. \textbf{As models continue to scale and improve, these strengths will likely grow even more pronounced, offering abundant opportunities to support large-scale video content analysis.}

Despite their potential, MLLMs can be misaligned with human perceptions. 

Our findings indicate that operationalizing abstract concepts with greater detail can enhance alignment. However, it may also risk constraining the MLLM’s ability to uncover novel social dynamics beyond the specified criteria. This contrasts with typical in-context or few-shot learning scenarios, where multiple demonstrations help the model infer task structure and reduce ambiguity by leveraging patterns recognized during pretraining \cite{min-etal-2022-rethinking}. \textbf{This challenge will likely persist regardless of model size since concepts are intrinsically ambiguous. Recent work like \cite{baluja-2025-text} and \cite{NEURIPS2024_540a6eef} demonstrate that even state-of-the-art MLLMs like GPT4 still face difficulties interpreting abstract concepts like humor in multimodal contexts such as memes, comics, or spoken conversations.} In diverse social media content, models must balance consistency with flexibility to adapt to dynamic contexts.

% Models must balance consistency with the flexibility to generalize across diverse and dynamic social media content. 
Additionally, when applying MLLM to analyze videos in the wild, we highlight that video style diversity is a crucial factor impacting model alignment. The short videos in our study are predominantly informal and casually filmed in everyday settings.
% often incorporating creative elements. 
They differ from vlogs, tutorials, streams, or product reviews, typically more structured and polished. Our findings show that the MLLM can struggle to capture and interpret unconventional visual cues, such as the novel yet subtle suggestion of suicide depicted in Figure \ref{fig:examples} (c). \textbf{Although more advanced MLLMs may be more generally aligned with human perceptions, these context-dependent and culturally specific signals often require situated awareness that larger-scale pre-training alone may not sufficiently address.} Developing and evaluating models that can effectively navigate such ambiguity while maintaining alignment on more structured formats remains essential for advancing multimodal analysis across diverse platforms.

\subsection{Limitations and Future Work}
% Based on our findings around the factors influencing human-AI alignment in concept annotations and understandings, 
% We propose the following future directions for better alignment in MLLM-assisted visual content analysis. 
% Based on the aforementioned discussion, 
We emphasize three directions to improve human-AI alignment in (M)LLM-assisted visual content analysis: human-centered auditing, multimodal synthesis, and temporality incorporation.

\textbf{Implementing MLLM response auditing.} In our case study, MLLM interpretations often diverged from human concept understanding due to factors like concept complexity and the diversity of video genres. Specifically, the MLLM may systematically misunderstand the visual cues of videos of specific genres, such as cartoons and memes, as suggested in Section~\ref{sec:Diverse}. Thus, it is crucial to implement human-centered post hoc audits \cite{xiao_human-centered_2024, zhang_redefining_2023}. Shen et al. \cite{shen2024valuecompass} developed a framework to audit the value alignment of humans and language models to improve transparency and ethical use of AI in social research.
Future work can explore incorporating human-centered evaluation as a standard step in MLLM-assisted content analysis workflows \cite{gero_supporting_2024, chen2024unifiedhallucinationdetectionmultimodal, leiser_hill_2024, yehuda-etal-2024-interrogatellm}. Such measures can facilitate the iterative refinement of concept operationalization and prompt engineering to address known biases in an AI's understanding of social concepts.

\textbf{Synthesizing multimodal inputs.}
In our current workflow, we decode videos into keyframes and prompt the MLLM to annotate concepts given isolated images. However, we can also incorporate audio or transcripts to provide a more comprehensive analysis, though interpreting signals from multiple input sources remains challenging. Additionally, as discussed in Section~\ref{sec:Diverse}, conflicting information across different modalities can complicate interpretations. Developing more sophisticated methods for synthesizing multimodal inputs is thus a promising avenue for future research.

\textbf{Incoporating video temporality.}
Some concepts require a temporal context for accurate interpretations. 
% MLLMs require more temporal contexts for visually dependent concepts, such as incorporating surrounding keyframes to better grasp dynamic content.
% such as the surrounding keyframes, which are limited in understanding the dynamic context in videos. 
For example, concepts like emotional valence and genre often depend on a holistic understanding of the video's overall narrative \cite{liu_harnessing_2024}, which isolated keyframes cannot capture. Future work could explore MLLMs that directly interpret videos or a sequence of keyframes to provide more contextual information. 

\subsection{Ethical Considerations}
Our findings suggest that human-AI misalignment may result in systematic biases. Previous studies have reported LLMs' biases towards minorities and underrepresented populations, including people with disabilities \cite{gadiraju_i_2023} and socially subordinate groups \cite{lee_large_2024}. Future studies can work on identifying the potential biases in LLMs.

\section{Conclusion}
We conduct one of the earliest case studies on leveraging Multimodal Large Language Models (MLLMs) to interpret abstract social concepts in video data. Whereas prior work primarily employed LLMs for text-based social media data, we demonstrate how MLLMs can extend large-scale automated content analysis to video content, capturing abstract concepts of self-disclosure styles and subjective visual cues. Through quantitative and qualitative comparisons, we highlight key factors undergirding misalignments between MLLM and human perceptions, such as concept operationalization, complexity, and genre diversity. Interestingly, adding prototypical manifestations of abstract concepts does not consistently improve alignment. Our results underscore the importance of post-hoc auditing and human oversight to ensure agreement between AI outputs and human understanding. Future work should explore the integration of multimodal inputs and experiment with fine-tuning or in-context learning to enhance the model’s ability to understand more complex social interactions.

\begin{acks}
    We appreciate the comments by Dr. Yunlong Wang. This project is partially supported by the University of Texas at Austin University Graduate Continuing Fellowship, Cisco, NSF grant IIS2107524, and Good Systems\footnote{\url{https://bridgingbarriers.utexas.edu/good-systems}} a UT Austin Grand Challenge to develop responsible AI technologies. The statements made herein are solely the opinions of the authors and do not reflect the views of the sponsoring agencies.
\end{acks}

\bibliographystyle{ACM-Reference-Format}
\bibliography{mentalhealth,otherbibs}
\appendix

\section{LLaVA Prompts}
\label{sec:LLaVAPrompts}
\onecolumn
\begin{longtable}{|p{0.10\textwidth}|p{0.10\textwidth}|p{0.70\textwidth}|}
\caption{LLaVA Prompts} \label{tab:evaluation} \\

\hline
\textbf{Concept} & \textbf{Strategy} & \textbf{Prompt} \\
\hline
\endfirsthead

\hline
\textbf{Concept} & \textbf{Strategy} & \textbf{Prompt} \\
\hline
\endhead

\hline
\endfoot

\hline
\endlastfoot

% \sam{put this in front of everything for footnotesize}
% \footnotesize 
Interacting &
  Prompt 0 - Naive &
  "\textless{}image\textgreater{}\newline USER: Does this picture portray an interacting style, yes or no? Explain your answer.\newline ASSISTANT:" \\

& Prompt 1 - Simple Definition &
  "\textless{}image\textgreater{}\newline USER: Interacting style refers to creators establishing a simulated interpersonal relationship with their audience, fostering a sense of engagement and connection. Does this picture portray an interacting style, yes or no? Explain your answer.\newline ASSISTANT:" \\

& Prompt 2 - Detailed Definition &
  "\textless{}image\textgreater{}\newline USER: Interacting style refers to creators establishing a simulated interpersonal relationship with their audience, fostering a sense of engagement and connection. This involves behaviors such as directly addressing the audience, using conversational language, or acknowledging comments or questions from viewers. Does this picture portray an interacting style, yes or no? Explain your answer.\newline ASSISTANT:" \\

& Prompt 3 - Openminded &
  "\textless{}image\textgreater{}\newline USER: Interacting style refers to creators establishing a simulated interpersonal relationship with their audience, fostering a sense of engagement and connection. This involves behaviors such as directly addressing the audience, using conversational language, or acknowledging comments or questions from viewers. These are just several examples, so be open-minded to other potential scenarios of interacting style. Does this picture portray an interacting style, yes or no? Explain your answer.\newline ASSISTANT:" \\

Presenting &
  Prompt 0 - Naive &
  "\textless{}image\textgreater{}\newline USER: Does this picture communicate in a presenting style, yes or no? Explain your answer.\newline ASSISTANT:" \\

& Prompt 1 - Simple Definition &
  "\textless{}image\textgreater{}\newline USER: Presenting style involves the delivery of information, typically accompanied by visual aids like slides or graphics. Does this picture communicate in a presenting style, yes or no? Explain your answer.\newline ASSISTANT:" \\

& Prompt 2 - Detailed Definition&
  "\textless{}image\textgreater{}\newline USER: Presenting style involves the delivery of information, typically accompanied by visual aids like slides or graphics, such as a businessman presenting slides, a student giving a speech on a topic, or a general rallying troops for war. Does this picture communicate in a presenting style, yes or no? Explain your answer.\newline ASSISTANT:" \\

& Prompt 3 - Openminded &
  "\textless{}image\textgreater{}\newline USER: Presenting style involves the delivery of information, typically accompanied by visual aids like slides or graphics, such as a businessman presenting slides, a student giving a speech on a topic, or a general rallying troops for war. These are just several examples, so be open-minded to other potential scenarios of presenting style. Does this picture communicate in a presenting style, yes or no? Explain your answer.\newline ASSISTANT:" \\

Arousal &
  Prompt 0 - Naive &
  "\textless{}image\textgreater{}\newline USER: What level of arousal does this image communicate, low, moderate, or high? Explain your answer.\newline ASSISTANT:" \\

& Prompt 1 - Simple Definition &
  "\textless{}image\textgreater{}\newline USER: Low arousal is associated with calmness, relaxation, or drowsiness. Moderate arousal is a balanced state of alertness and engagement without overstimulation. High arousal is characterized by heightened physiological and emotional activity. What level of arousal does this image communicate, low, moderate, or high? Explain your answer.\newline ASSISTANT:" \\
 &
  Prompt 2 - Detailed &
  "\textless{}image\textgreater{}\newline USER: Low arousal is associated with calmness, relaxation, or drowsiness. For example, feeling fatigued or viewing a peaceful landscape or a calm, monochromatic image. Moderate arousal is a balanced state of alertness and engagement without overstimulation, often linked with optimal performance and involves minimal physiological activation. For example, feeling attentive or focused; engaging in a conversation or viewing a moderately complex image. High arousal is characterized by heightened physiological and emotional activity. For example, feeling excited, anxious, or stressed; or viewing a dynamic or chaotic scene with bright colors or intense stimuli. What level of arousal does this image communicate, low, moderate, or high? Explain your answer? \textbackslash{}nASSISTANT:" \\
 &
  Prompt 3 - Open Minded &
  "\textless{}image\textgreater{} \newline USER: Low arousal is associated with calmness, relaxation, or drowsiness. For example, feeling fatigued or viewing a peaceful landscape or a calm, monochromatic image. Moderate arousal is a balanced state of alertness and engagement without overstimulation, often linked with optimal performance and involves minimal physiological activation. For example, feeling attentive or focused; engaging in a conversation or viewing a moderately complex image. High arousal is characterized by heightened physiological and emotional activity. For example, feeling excited, anxious, or stressed; or viewing a dynamic or chaotic scene with bright colors or intense stimuli. These are just several examples so be open-minded to other potential scenarios of arousal levels. What level of arousal does this image communicate, low, moderate, or high? Explain your answer? \textbackslash{}nASSISTANT:" \\
Diversity &
  Prompt 0 - Naive &
  "\textless{}image\textgreater{}\newline USER: What level of diversity does this image communicate, low, moderate, or high? Explain your answer? \textbackslash{}nASSISTANT:" \\
 &
  Prompt 1 - Definition &
  "\textless{}image\textgreater{}\newline USER: The diversity of an image includes the color variation, compositional complexity, and originality of the image. What level of diversity does this image communicate, low, moderate, or high? Explain your answer? \textbackslash{}nASSISTANT:" \\
 &
  Prompt 2 - Detailed &
  "\textless{}image\textgreater{}\newline USER: The diversity of an image includes the color variation, compositional complexity, and originality of the image. Color variation involves assessing the range of colors across the image. Compositional complexity involves the arrangement of diverse elements within the image. Originality assesses whether the image presents a new or uncommon perspective. What level of diversity does this image communicate, low, moderate, or high? Explain your answer? \textbackslash{}nASSISTANT:" \\
 &
  Prompt 3 - Open Minded &
  "\textless{}image\textgreater{}\newline USER: The diversity of an image includes the color variation, compositional complexity, and originality of the image. Color variation involves assessing the range of colors across the image. Compositional complexity involves the arrangement of diverse elements within the image. Originality assesses whether the image presents a new or uncommon perspective. These are just several examples so be open-minded to other instances of diversity. What level of diversity does this image communicate, low, moderate, or high? Explain your answer? \textbackslash{}nASSISTANT:"

\end{longtable}
\twocolumn

\section{Bootstraping Details}
\label{sec:bootstrap}

To assess how human-AI alignment differs across configurations for each concept, we employ a bootstrapping approach inspired by the methodology outlined in \cite{berg-kirkpatrick-etal-2012-empirical}. We first collect a pool of generated annotations for each concept and prompt configuration to compute an initial alignment score. However, relying on a single measure fails to capture the variability inherent in the data, and observed differences across configurations may arise purely by chance. This limitation makes it challenging to draw reliable conclusions about the relative alignment of different prompt configurations. 

The bootstrapping approach addresses this issue by repeatedly resampling the data to estimate the variability in alignment scores. Specifically, we generate $N$ resampled datasets, each of size $K$, by randomly drawing annotations with replacement from the original pool. An alignment score is computed for each resampled dataset, resulting in a distribution of $N$ scores for each concept-prompt pair. This distribution reflects the variability in alignment and enables us to assess, on average, how reliably each prompting configuration aligns with human perceptions across the selected social concepts beyond random chance. We visualize these score distributions in Figure~\ref{fig:Density}) and discuss findings in Section~\ref{sec:RQ1}.

\section{Ethics Statement}
\label{sec:ethics}

We are committed to conducting ethically responsible research, ensuring content creators' privacy, and safeguarding research team members' well-being. Since this study analyzes data on publicly available platforms like YouTube, it qualifies for human subjects exemption under our university's Institutional Review Board (IRB) guidelines, posing minimal risk to content creators (or individuals present in the video). Nevertheless, we acknowledge that creators could not provide explicit consent for the inclusion or exclusion of their content. To respect the creator's privacy, we implemented additional protections, such as anonymizing individuals in the video by obscuring facial features in any snapshots in this paper. Additionally, we do not collect personally identifiable metadata about the creators or individuals presented in the videos. 

Another ethical factor is the well-being of researchers exposed to potentially distressing material, particularly during qualitative analyses involving sensitive topics like depression. To mitigate potential emotional harm, we provided team members access to university mental health resources, encouraged breaks during data analysis, and fostered an environment of open communication about the work's emotional impact.
\end{document}